\documentclass[superscriptaddress,amsmath,amssymb,aps,prl,twocolumn]{revtex4-1}

\usepackage{epsfig}
\usepackage{epstopdf}
\usepackage{graphicx}
\usepackage{dcolumn}
\usepackage{bm}
\usepackage{amsmath}
\usepackage{array}
\usepackage{color}

\begin{document}

\setlength{\belowdisplayskip}{0pt} \setlength{\belowdisplayshortskip}{0pt}
\setlength{\abovedisplayskip}{0pt} \setlength{\abovedisplayshortskip}{0pt}

\title{Nanophotonic coherent light-matter interfaces based on rare-earth-doped crystals}

\author{Tian Zhong}
\affiliation{T. J. Watson Laboratory of Applied Physics, California Institute of Technology, 1200 E California Blvd, Pasadena, CA, 91125, USA}

\author{Jonathan M. Kindem}
\affiliation{T. J. Watson Laboratory of Applied Physics, California Institute of Technology, 1200 E California Blvd, Pasadena, CA, 91125, USA}

\author{Evan Miyazono}
\affiliation{T. J. Watson Laboratory of Applied Physics, California Institute of Technology, 1200 E California Blvd, Pasadena, CA, 91125, USA}

\author{Andrei Faraon}
\email{faraon@caltech.edu}
\affiliation{T. J. Watson Laboratory of Applied Physics, California Institute of Technology, 1200 E California Blvd, Pasadena, CA, 91125, USA}

\date{\today}

\maketitle


\noindent \textbf{Quantum light-matter interfaces (QLMIs) connecting stationary qubits to photons will enable optical networks for quantum communications, precise global time keeping, photon switching, and studies of fundamental physics. Rare-earth-ion (REI) doped crystals are state-of-the-art materials for optical quantum memories and quantum transducers between optical photons, microwave photons and spin waves. Here we demonstrate coupling of an ensemble of neodymium REIs to photonic nano-cavities fabricated in the yttrium orthosilicate host crystal. Cavity quantum electrodynamics effects including Purcell enhancement (F=42) and dipole-induced transparency are observed on the highly coherent $^4I_{9/2}-^4F_{3/2}$ optical transition. Fluctuations in the cavity transmission due to statistical fine structure of the atomic density are measured, indicating operation at the quantum level. Coherent optical control of cavity-coupled REIs is performed via photon echoes. Long optical coherence times ($T_2 \sim$100 $\mu s$) and small inhomogeneous broadening are measured for the cavity-coupled REIs, thus demonstrating their potential for on-chip scalable QLMIs.}

QLMIs are quantum devices composed of light emitters with quantum states that can be controlled via optical fields and entangled to photons~\cite{Kimble, Komar}. They enable distribution of quantum entanglement over long distances in optical quantum networks for quantum communications~\cite{Kimble}. Quantum networks of atomic clocks have also been proposed for precise global time-keeping and studies of fundamental physics~\cite{Komar}. Realizing QLMIs requires control of light and matter at the single atom and single photon level, which enable optoelectronic devices like optical modulators and nonlinear optical devices operating at the most fundamental level~\cite{Chen}. QLMIs are also expected to play a leading role in realizing optical to microwave quantum transducers for interconnecting future superconducting quantum machines via optical fibres~\cite{Williamson, Andrews}. 

Scalable and robust QLMIs require emitters to have long spin coherence times and coherent optical transitions. For integrated optical quantum networks, these emitters need to be coupled to on-chip optical resonators that capture the photons in a single mode and further couple them into optical fibres or waveguides. The solid-state emitters most investigated so far for on-chip QLMIs are semiconductor quantum dots (QDs)~\cite{Michler} and nitrogen vacancy centers in diamond (NVs)~\cite{Igor}. To date, complete quantum control of single QD and NV spins, spin-photon entanglement, and entanglement of remote NVs via photons have been realized~\cite{Press, Greve, Bernien}. Both QDs~\cite{Englund} and NVs~\cite{Faraon} have been coupled to optical nano-cavities. However, the challenge in growing optically identical QDs limits their prospects for a scalable architecture~\cite{Michler}. NVs embedded in nanostructures have long electronic spin coherence times~\cite{Li}, but suffer from optical spectral instabilities such as blinking and spectral diffusion~\cite{Faraon12}. These spectral instabilities have so far impeded the coherent coupling between optical fields and NV centers in nano-resonators that are essential for further developments of QLMIs. 

Rare-earth ions (REIs) embedded in host crystals at cryogenic temperatures exhibit highly coherent quantum states in the 4f orbital~\cite{Thiel}. The Zeeman or hyperfine states of REIs can have coherence times as long as six hours~\cite{Zhong}, the longest ever demonstrated in a solid. These states are connected via optical transitions with the narrowest linewidth in the solid state (sub-kHz) and small inhomogeneous broadening (MHz to GHz)~\cite{Sun}. This outstanding optical and spin coherence makes REI-doped crystals the state-of-the-art material for macroscopic solid-state optical quantum memories~\cite{Lvovsky, Tittel}. Integrated REI-doped waveguide quantum memories have also been developed~\cite{Erhan, Marzban}. Detection and control of single REI spins has been recently demonstrated in bulk material, but not using the transitions employed in optical quantum memories at cryogenic temperatures~\cite{KolesovNC, Utikal}. Coupling the highly coherent optical transitions of REIs to nano-cavities will enable on-chip QLMIs where REI ensembles act as quantum memories and single REIs act as qubits~\cite{Mcauslan}.

Here we demonstrate high-cooperativity coupling of a neodymium (Nd$^{3+}$) ensemble to photonic nano-cavities fabricated directly in the yttrium orthosilicate (YSO) host crystal, and show coherent optical control of REIs coupled to nanophotonic cavities. These results are enabled by the long coherence time and small inhomogeneous broadening of cavity-coupled REIs, which are essential properties that may lead to nanophotonic QLMIs with better prospects for scalability than those based on NVs and QDs.

  \begin{figure*}[htb]
\includegraphics[width=1\textwidth]{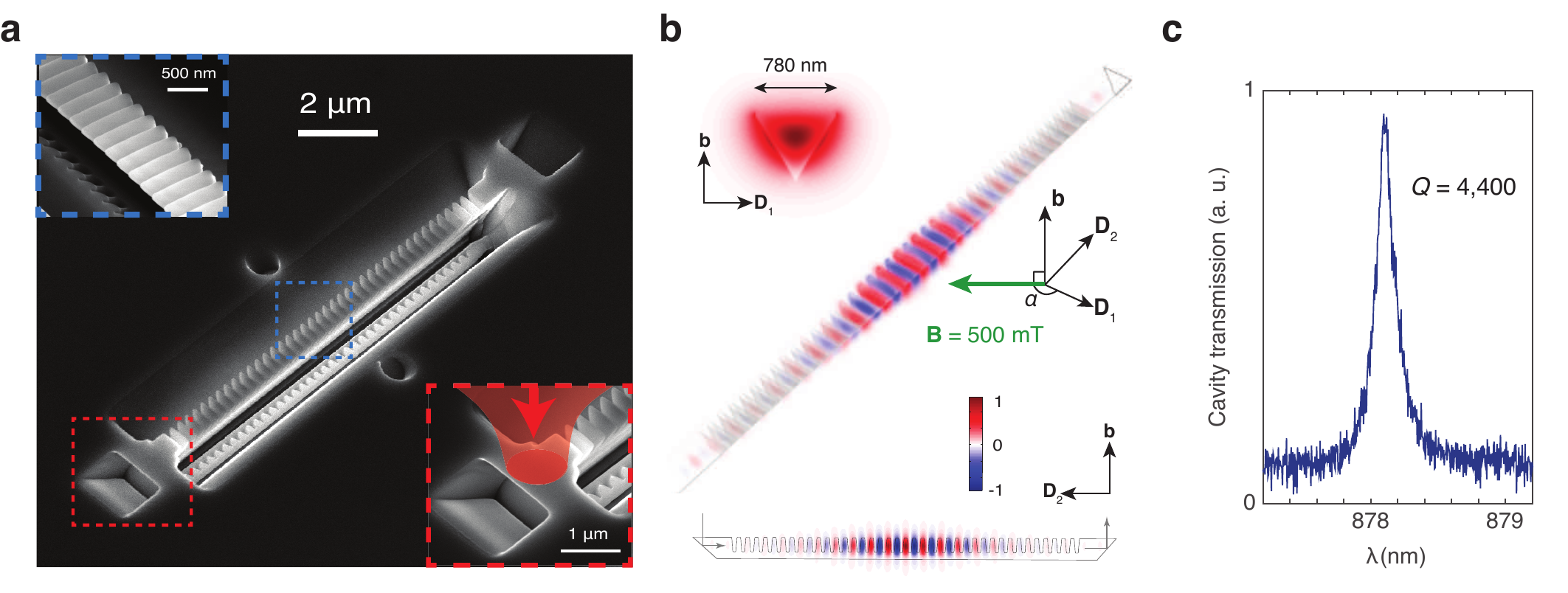}
\caption{{\bf Photonic crystal nano-beam resonator fabricated in Nd:YSO.} {\bf a.} Scanning electron microscope image of the device. The red inset is a zoomed-in view of the 45$^\circ$ angle-cut coupler that allows vertical coupling of light from a microscope objective. The blue inset shows the grooves forming the photonic crystal. {\bf b.} Schematics of the nano-beam resonator with simulated field profiles of the fundamental TE resonance mode. The TE polarization aligns with the D$_1$ axis of the YSO crystal. A magnetic field of 500 mT is applied in the D$_1$-D$_2$ plane at an angle of $\alpha$=135$^{\circ}$ with respect to D$_1$ axis. {\bf c.} Broadband cavity transmission spectrum showing the cavity resonance with quality factor $Q$=4,400.}
\label{f1}
\end{figure*}

 \section*{Results}

\noindent {\bf Design, fabrication and characterization of nano-cavities.} The nano-cavities, one of which is shown in Fig.~\ref{f1}(a), were fabricated in neodymium-doped yttrium orthosilicate (Nd$^{3+}$:YSO) using focused ion beam milling. For this study, we used devices fabricated in two types of samples with Nd doping of 0.2\% and 0.003\% (Scientific Materials Inc.). The photonic crystal cavity is made of grooves milled in a triangular nano-beam~\cite{Bayn} (Fig.~\ref{f1}(b)) (see Methods). Finite-difference time-domain (FDTD) simulations~\cite{MEEP} indicate a TE mode with quality factor exceeding 1$\times10^5$, mode volume  $V_{\rm mode}=1.65(\lambda/n_{\rm YSO})^3=0.2\mu$m$^3$, and mode profile shown in Fig.~\ref{f1}(b). Here $V$ is defined as $V_{\rm mode}=\int_V \epsilon(\rm r)|\rm E(r)|^2d^3\rm r /\rm max (\epsilon(\rm r)|\rm E(r)|^2)$, where $\rm E(\rm r)$ is the electric field and $\epsilon(\rm r)$ is the electric permittivity at position $\rm r$. Two 45$^\circ$ angled cuts at both ends of the nano-beam (i.e. couplers) allow for coupling light from a direction normal to the chip (i.e. $\vec{\rm b}$ in Fig.~\ref{f1}(b)) using a confocal microscope setup (See Methods). A broadband light source was coupled into the resonator from one end and the transmitted light was collected from the other coupler with typical efficiencies ranging from 20\% to 50\%. The transmitted spectrum shows a resonance with quality factor $Q$=4,400 (Fig.~\ref{f1}(c)) in the device used for the following measurements. Arrays of devices were reproducibly fabricated with similar performance (Supplementary S1).

\begin{figure*}[htb]
\includegraphics[width=1\textwidth]{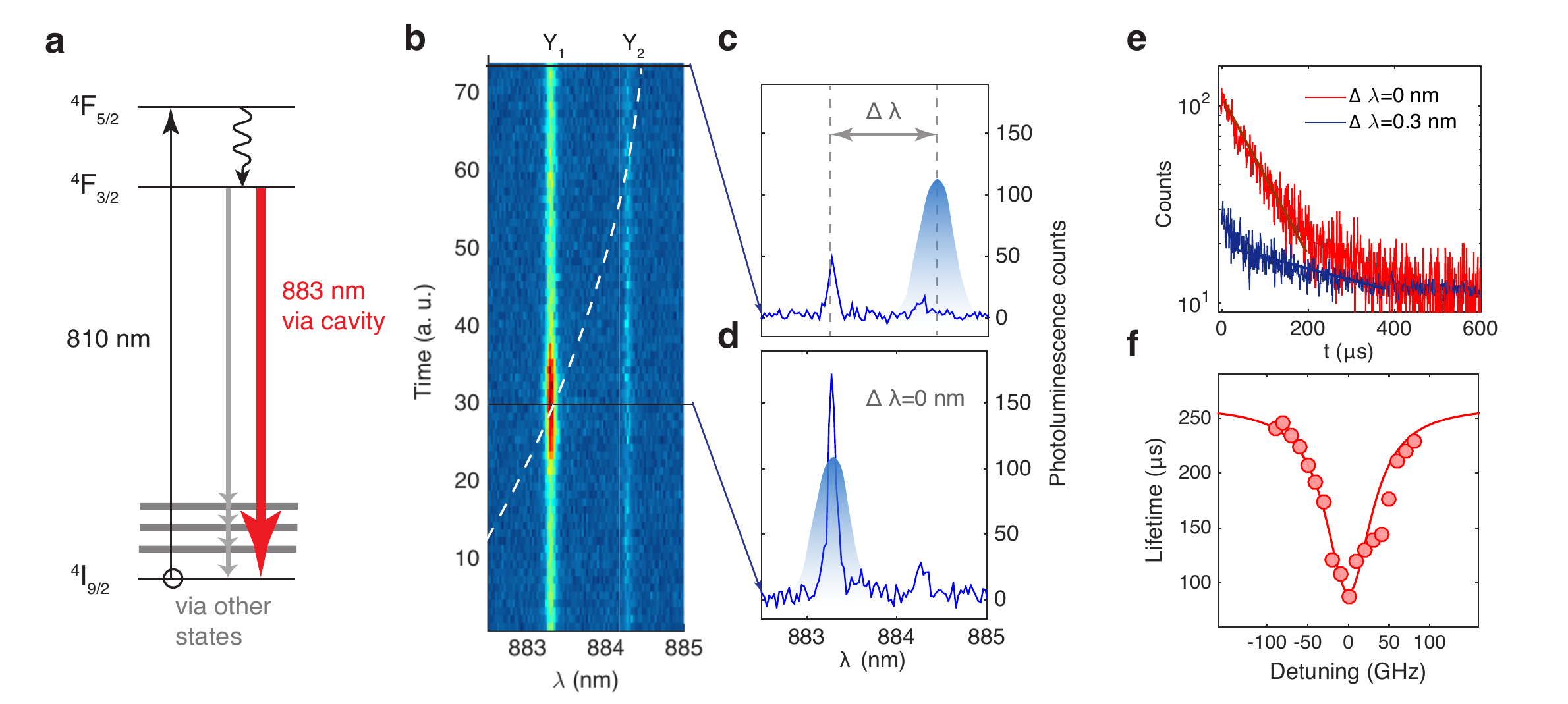}
\caption{{\bf Coupling of Nd$^{3+}$ ions to the YSO cavity mode showing enhanced photoluminescence (PL) and reduced lifetimes.} {\bf a.} Schematic of energy levels for Nd$^{3+}$ in YSO. Optical excitation at 810 nm results in PL at several wavelengths with only the 883 nm transition enhanced by the cavity. {\bf b.} Spectrogram showing the Nd$^{3+}$ PL while the cavity is tuned across resonance using gas condensation. The dashed line is a guide to the eye indicating the central wavelength of the cavity resonance. The cavity resonance is not visible because there is no background luminescence to populate the cavity mode. {\bf c. d.} PL spectra in the uncoupled ({\bf c}) and coupled ({\bf d}) cases. The cavity resonance was drawn to indicate the cavity location. {\bf e.} Lifetime measurements for coupled ($\tau^{c}$=87 $\mu$s, $\Delta\lambda=0$) and uncoupled ($\tau^{0}$=254 $\mu$s, $\Delta\lambda$=0.3 nm) cases. {\bf f.} Change in lifetime as a function of the cavity detuning, which fits well with the calculation (red curve) using quality factor $Q$=4,400, 4.5\% branching ratio and field intensity averaged over the mode volume. }
\label{f2}
\end{figure*}

\noindent {\bf Measurement of the coupling rate between REIs and the nano-cavity.} The coupling of Nd$^{3+}$ ions to the nano-cavity was observed through enhancement in photoluminescence (PL) and emission rates. With the 0.2\% device cooled at 3.5 K, an 810 nm laser coupled into the cavity excited PL in the $^4I_{9/2}-^4F_{3/2}$ transition that was then collected from the output coupler (Fig.~\ref{f2}(a)). The PL spectrum shows two lines at 883.05 nm and 884.06 nm, corresponding to two inequivalent sites (Y$_1$ and Y$_2$) of Nd$^{3+}$ in YSO. An important observation is that the inhomogeneous linewidth of the ions in the cavity is the same as in the bulk material, for both the 0.2\% ($\Delta_{\rm inhom}$=16.0 GHz) and 0.003\% (5.9 GHz) devices (Supplementary S2). A small inhomogeneous linewidth (on the order of $\sim10$ GHz) is important for scaling to networks of multiple QLMIs (Supplementary S3). The cavity resonance was tuned across the Nd$^{3+}$ PL line using a gas condensation technique~\cite{Faraon}. The spectrograms in Fig.~\ref{f2}(b-d) show enhancement of the Y$_1$ line when the cavity is resonant with it, a signature of coupling. The Y$_2$ line exhibits negligible enhancement because its dipole moment is not aligned with the TE cavity polarization (${\rm D}_1$ axis of the YSO crystal (Fig.~\ref{f1}(b)). The spontaneous emission rate enhancement was characterized via lifetime measurements. A pulsed laser at 810 nm excited fluorescence of the Y$_1$ line, which was filtered using a monochromator and detected with a single photon counter (Fig.~\ref{f2}(e)).  From single exponential fits, we calculated a reduction in lifetime from 254 $\mu s$ when the cavity is detuned by $\Delta \lambda=$0.3 nm, to 87 $\mu s$ on resonance. Taking into account the branching ratio of the 883 nm transition ($\beta\sim4.5\%$, see Methods), the reduction in lifetimes corresponds to an ensemble averaged Purcell factor~\cite{Purcell} F=42, which agrees well with the estimations that assume a uniform spatial distribution of Nd$^{3+}$ ions in the resonator (Supplementary S4). A single ion positioned at the maximum cavity field would experience a Purcell factor $\sim$200. A similar result was obtained in a 0.003\% cavity with lower quality factor (Supplementary S5).

\begin{figure*}[ht]
\includegraphics[width=.9\textwidth]{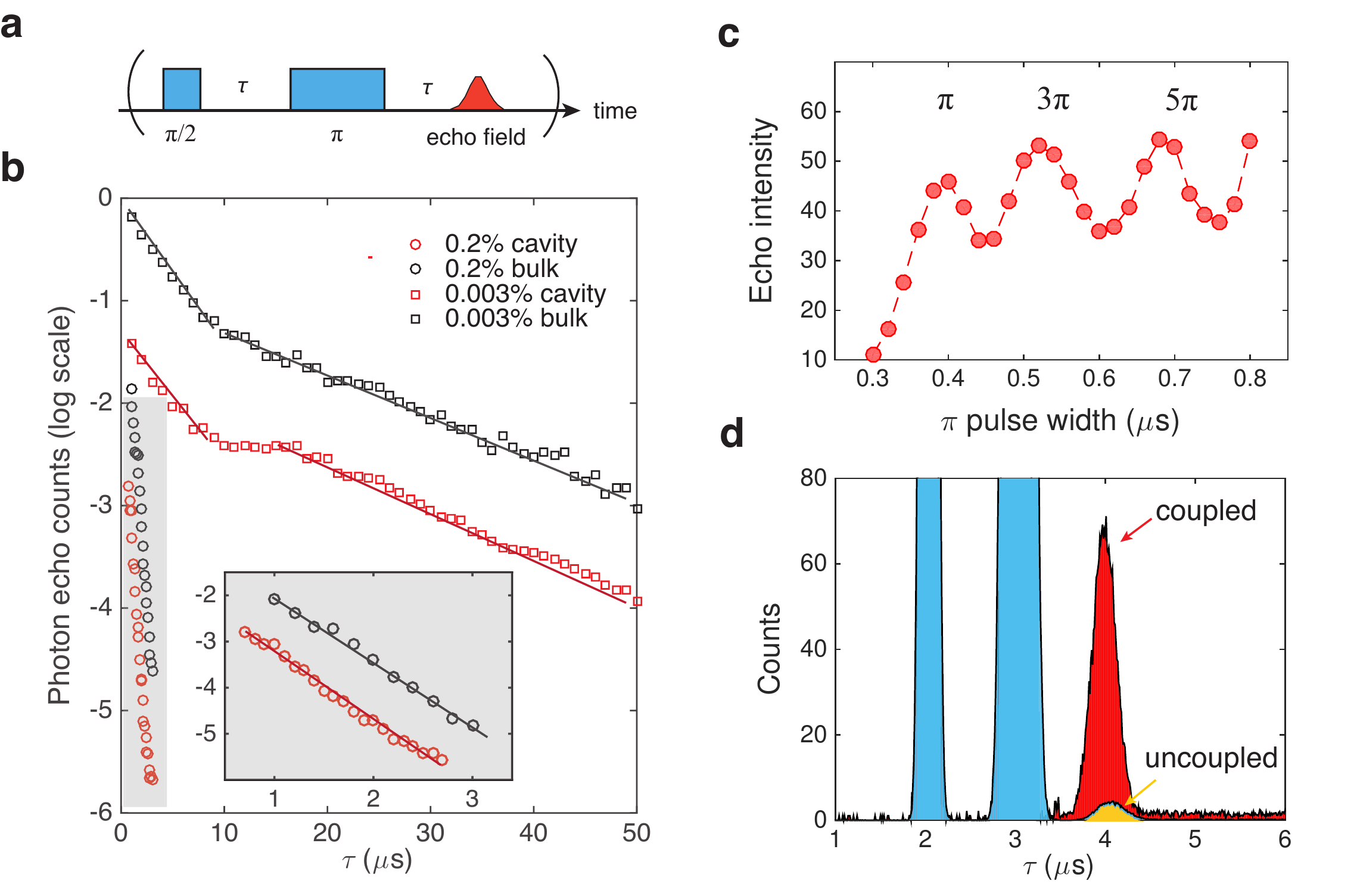}
\caption{{\bf Photon echo measurements showing long optical coherence time and enhanced echo intensity from an ensemble of Nd$^{3+}$ ions in the cavity.} {\bf a.} Two-pulse photon echo sequence ($\pi/2-\pi$) used to measure $T_2$. {\bf b.} Two-pulse photon echo decays measured in both the cavity (red) and the bulk (black) samples with two different doping concentrations. The inset plots the echo decays measured with a 0.2\% doped sample. {\bf c.} Oscillation of echo intensity with increasing width of the $\pi$ rephasing pulse. The periodic signal reveals the ensemble averaged Rabi frequency of the coupled ions. The ideal $\pi$ pulse duration for the input power was 0.4 $\mu$s. {\bf d.} Enhanced photon echo intensity (by $\sim$12 fold) when the cavity is coupled, compared to the uncoupled case (cavity detuned by $\Delta \lambda$=15 nm so that the transition is outside the photonic bandgap).}
\label{f3}
\end{figure*}

\noindent {\bf Measurement of the optical coherence time for cavity-coupled REIs.} Coherent and stable optical transitions are essential for QLMIs. We characterized the optical coherence time $T_2$ of the 883 nm transition using two-pulse ($\pi/2$ - $\pi$) photon echo techniques (Fig.~\ref{f3}(a)), with an applied magnetic field of B=0.5 T (see Methods). The laser pulses were coupled in and the echoes were collected via the couplers while the 0.2\% and 0.003\% cavities were on resonance with the Nd transition.  Since only a small sub-ensemble ($<$100 ions) in the cavity was excited, the weak echo signal required detection using single photon counters. A typical echo from the 0.2\% cavity is shown in Fig.~\ref{f3}(d). The echo decays as a function of the ($\pi/2$ - $\pi$) time delay $\tau$ are plotted in Fig.~\ref{f3}(b) together with photon echoes from bulk substrates. For the 0.2\% sample, $T^{c}_2$=4$\tau^c$=2.8$\pm$0.4 $\mu$s was measured for the cavity, which shows a good agreement with the bulk value of $T^{b}_2$=3.2$\pm$0.4 $\mu$s. For the 0.003\% sample, the echo exhibited two exponential decays. The slower decays give $T^{b}_2$=100$\pm$5$\mu$s (bulk) and $T^{c}_2$=94$\pm$5 $\mu$s (cavity), which match with values reported in~\cite{Usmani}.  The fast decays are likely due to the superhyperfine interactions between Nd$^{3+}$ and its neighbouring yttrium ions, which commonly manifests as modulated echoes decaying faster than $T_2$ \cite{Usmani}. No oscillations was observed in Fig.~\ref{f3}(a) because of the fast modulation frequency ($\sim$1 MHz) due to our strong magnetic field. By lowering the B field, we observed oscillations of echoes for the initial 10 $\mu$s delay. Additionally, we varied the excitation powers and did not see changes in the $T_2$ values, which indicates our measurement was not significantly affected by instantaneous spectral diffusion. In sum, the good agreement on $T_2$ between the cavity and bulk confirms that the optical coherence property of Nd$^{3+}$ ions is not affected by the nano-fabrication. For higher Purcell factors, the $T_2$ in cavities should decrease due to the $T_2\leq2T_1$ limit and would become smaller than the bulk value. This regime is not reached here because the Purcell enhanced $2T_1$ is not smaller than $T_2^b$.

The observation of photon echoes demonstrates coherent optical control of the quantum state of cavity-coupled ions. This control was further extended by varying the $\pi$ pulse duration and observing Rabi oscillations in the echo intensities as shown in Fig.~\ref{f3}(c). A Rabi frequency of $\sim$6 MHz is estimated. The same oscillation was not observed in the bulk. For the coupled laser power, the optimal $\pi$ pulse duration is 0.4 $\mu$s. The oscillations are not visible for pulse duration less than 0.3 $\mu$s, because of the limited rise/fall times ($\sim200$ ns) of the pulse-generating setup (Methods).  A $\sim$12 fold increase in the echo intensity is observed in the cavity-coupled case compared to the uncoupled case (i.e. detuning $\Delta \lambda$=15 nm) as shown in Fig.~\ref{f3}(d). This enhancement can be attributed to a combination of several effects: the higher atomic absorption rate through the Purcell effect~\cite{AfzeliusIM}, stronger intra-cavity field intensity, and high echo collection efficiency as the ions emit dominantly into the cavity mode. The spectral diffusion of the coupled ions using three-pulse photon echoes was also investigated. The homogeneous linewidths were broadened at rates of 6.1 kHz $\mu$s$^{-1}$ for the 0.2\% doped cavity and 380 Hz $\mu$s$^{-1}$ for the 0.003\% cavity. These slow spectral diffusions permit repeated optical addressing of the ions for 10s of $\mu$s (Supplementary S6).

\begin{figure*}[htb]
\includegraphics[width=1\textwidth]{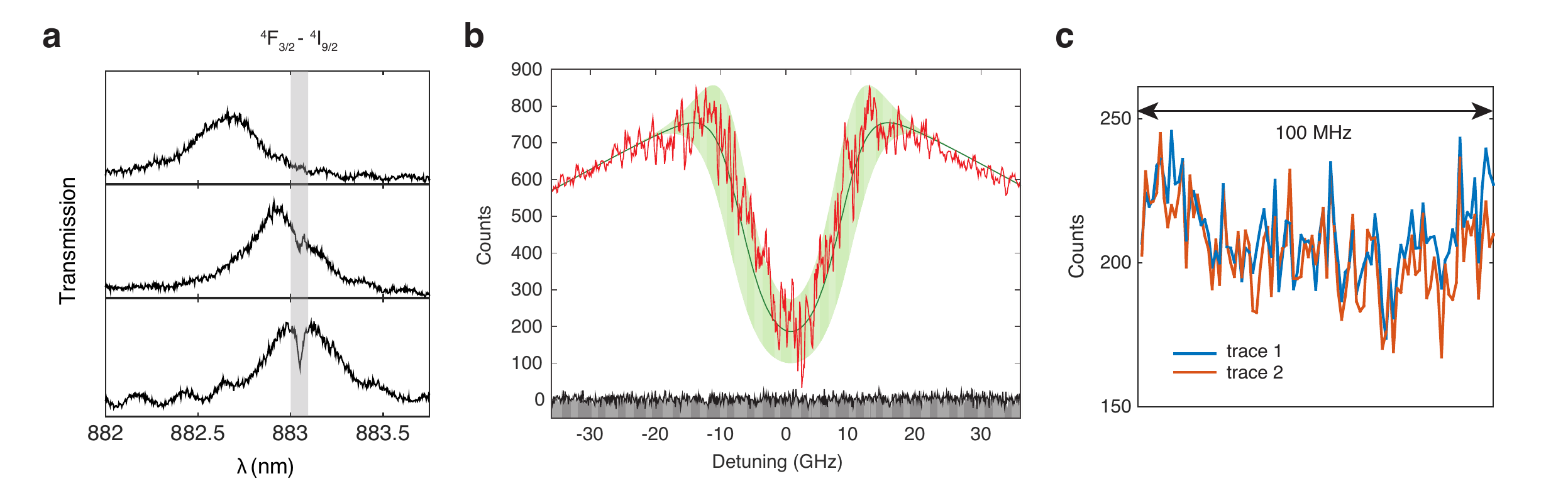}
\caption{{\bf Control of cavity transmission and observation of statistical fine structure (SFS) of coherently driven Nd$^{3+}$ ions in the cavity.} {\bf a.} Broadband transmission spectra as the cavity is tuned to the 883 nm Nd transition. A dip is observed when the two are on resonance. The negligible dip at far detunings confirms that this effect is not due to absorption, but quantum interference between the intra-cavity field and the ions. {\bf b.} High resolution transmission spectrum (red curve) obtained by scanning a narrow linewidth ($\sim$20 KHz) Ti:Sapphire laser over the inhomogeneous line. Green curve is the fit using parameters: $\bar{g}$=2$\pi\times$6 MHz, $\Gamma_{\rm h}$=100 kHz, $\Gamma_{\rm inhom}$=16.0 GHz, and assuming a Gaussian ion density distribution. The green shaded region is the estimated fluctuation in the transmitted laser intensity caused by $\sqrt{N}$ statistical variations of the ion density. Large fluctuations are expected because the density $N$ is low (few tens), which agrees with the measurement. The fluctuations within the inhomogeneous linewidth is noticeably larger than that at far detunings ($>$25 GHz) and the technical background noise (grey area), confirming that they are caused by SFS of the ion spectral density. {\bf c.}  Two traces of the transmitted intensities over the same 100 MHz bandwidth near zero detuning at different times. The high degree of correlation confirms the static and repeatable nature of SFS.}
\label{f4}
\end{figure*}

\noindent  {\bf Measurement of dipole-induced transparency and statistical fine structure.} QLMIs require efficient interactions between atoms and photons, which is why quantum memories use long atomic clouds or doped crystals to achieve large optical depth. One key advantage provided by nano-resonators is that efficient atom-photon interaction can be achieved in a small volume with only a handful of ions. This is readily observable in our system, where the Nd$^{3+}$ ions coherently interact with the intra-cavity field and control its transmission via dipole-induced transparency~\cite{Waks}. With the cavity tuned to 883 nm, the cavity transmission was probed using broadband light and a dip was observed at resonance (Fig.~\ref{f4}(a)). The depth of the dip depends on the collective coupling cooperativity $\eta=4N\bar{g}^2/(\kappa\Gamma_{\rm h})$, where $\bar{g}$ is the ensemble averaged coupling strength, $\kappa$ is the cavity full linewidth,  $\Gamma_{\rm h}$ is the Nd$^{3+}$ homogeneous linewidth, and $N$ is the number of ions per $\Gamma_{\rm h}$. Considering an empirical collective dipole-cavity coupling model, the normalized cavity transmission in the presence of unsaturated resonant ions is,

\begin{equation}
T=\left|\frac{\kappa}{i\Delta+\kappa+4N\bar{g}^2/\gamma_{\rm h}}\right|^{2},
\end{equation}

\noindent which simplifies to $T=(1+\eta)^{-2}$ for zero detuning~\cite{Thompson}. The cavity transmission can be controlled by varying the probe light power and observing the saturation of the ions at increasing intra-cavity photon number (Supplementary S7). The saturation photon number in the nano-cavity was measured to be $\langle n_{\rm cav} \rangle=2\times10^{-5}$.

To better resolve the spectrum, a narrow ($\sim$20 kHz) Ti:Sapphire laser was scanned across the resonance (See Methods) to give the transmitted signals shown in Fig.~\ref{f4}(b, c). A 75 \% decrease in transmission was measured at zero detuning, which corresponds to a collective cooperativity $\eta\sim1.2$. Fitting using a Gaussian spectral density distribution (green line) with measured parameters $\bar{g}$=2$\pi\times$6 MHz, $\Gamma_{\rm h}$=100 kHz, $\Gamma_{\rm inhom}$=16.0 GHz, gives a peak ion density of $N\approx$53. Due to the statistical fine structure (SFS) of the inhomogeneously broadened line, a variation in the transmitted intensity ($\langle \delta T \rangle^2\sim\frac{d^2T}{dN^2}\langle \delta N\rangle^2$) due to $\delta N=\sqrt{N}$ fluctuations in the ion spectral density is expected~\cite{Moerner}. This expected variation is represented in Fig.~\ref{f4}(b) by the green shaded region and shows good agreement with that of the measured signal. This variation within the inhomogeneously broadened line, on which the statistical fluctuations of the ion spectral density are imprinted, is significantly larger than the spectrometer technical noise (grey area) and laser shot noise at far detunings ($>$25GHz), thus confirming that the static SFS in spectral density $N$($\Delta \lambda$) is probed. Two traces of the laser scan over the same 100 MHz bandwidth near zero detuning at different times are shown in Fig.~\ref{f4}(c). The high degree of correlation reflects the static and repeatable nature of SFS. Notably, the current platform would allow detection and control of a single ion coupled to the cavity if $N<<$1 and the laser linewidth were narrower than $\Gamma_h$ (Supplementary S8).

 \section*{Discussion}
\noindent The results reported in this paper (long optical coherence time, small inhomogeneous broadening, enhanced coherent optical control, and resonant probing of cavity-coupled REIs) demonstrate REI-based nanophotonics as a promising approach for robust and scalable quantum photonic networks integrating memories and single REI qubits. Single photon rates exceeding 1 MHz can be achieved with single REIs in nano-cavities with $Q$/$V\sim$10$^4$-10$^5$ ($V$ is normalized to $(\lambda/n)^3$), and the inhomogeneous broadening allows for frequency multiplexing of multiple REIs.  To use the interface as an optical quantum memory, efficient optical pumping into the long-lived Zeeman level needs to be demonstrated. Bulk REI quantum memories already boast high storage efficiency \cite{Hedges} with multi-mode capacity~\cite{Usmani}. Their implementations in our nanophotonic platform open the possibility of multiplexed systems for on-chip quantum repeaters. For Nd, high-fidelity storage of entanglement based on atomic frequency comb (AFC) has been demonstrated \cite{Clausen, Bussieres}. With cavity impedance matching \cite{AfzeliusIM}, unit storage efficiency is achievable with a mesoscopic ensemble of cavity-coupled ions. Meanwhile, long-lived nuclear spin coherence of 9 ms in $^{145}$Nd \cite{Wolfowicz} bodes well for spin-wave quantum memories using our nano-resonators. These devices can be further coupled to superconducting or optomechanical devices to enable hybrid quantum systems \cite{Williamson}. Furthermore, the technology can be readily transferred to other wavelengths, such as 1.5 $\mu$m for telecom quantum memories using Er$^{3+}$:YSO or 590 nm for long-haul quantum hard drives using Eu$^{3+}$:YSO \cite{Zhong}.

\section*{Methods}
\noindent {\bf YSO nano-resonator design and fabrication.}
The nano-beam has an equilateral triangular cross-section with each side of 780 nm. This geometry allows a circular fundamental mode field that can be efficiently coupled with a free space laser beam. The cavity is formed by forty equally spaced grooves of lattice constant 340 nm on the nano-beam, except for a defect introduced at the center by perturbing the lattice constant. The depth of the grooves is 65\% of the beam height. The triangular nano-beam resonator was fabricated using focused ion beam (FIB) milling followed by wet etching of Ga+ contaminated YSO in diluted (10\%) hydrochroric acid. An ion beam of 20 kV, 0.2 nA was used to fabricate the suspended nano-beam waveguide by milling at 30$^{\circ}$ angle with respect to the crystal surface normal. A small ion beam of 23 pA was then used to accurately pattern the grooves on top of the nano-beam. Limited by the finite width of the FIB beam, the side-walls of the grooves in the actual device were not vertical, but had an angle of 6$^{\circ}$. This leads to a degraded theoretical $Q$ of $5.0\times10^4$. We were able to reproducibly fabricate arrays of resonators (up to six) in a batch (Supplementary Information S1), with all the devices measuring resonances close to 883 nm and quality factors varying from 1,100 to 10,000.

\noindent {\bf Experimental setup for the photon echo measurements.}
A 500 mT external magnetic field was applied at $\alpha$=135$^{\circ}$ relative to the crystal D$_1$ axis using a pair of permanent magnets (see Fig.~\ref{f1}(b)). The $\pi/2$ and $\pi$ Gaussian pulses were generated by amplitude-modulating the Ti:Sapphire laser with two acousto-optic modulators (AOM) in series, with each in a double-pass configuration. The two pulse widths were 250 ns and 400 ns at a repetition rate of 1 kHz. { The average (peak) power of the excitation pulses was 210 nW (320 $\mu$W) measured after the objective lens. The extinction ratio between the on and off level of the pulses was $\sim$120 dB, ensuring sufficient signal-to-noise ratio for detecting echo photons using a Si single-photon counter (Perkin Elmer SPCM).  A third shutter AOM in single-pass configuration was inserted just before the photon counter to block the strong excitation pulses from saturating the detector. The extinction ratio of this shutter AOM was 30 dB.

\noindent {\bf High-resolution laser spectroscopy on cavity-coupled Nd$^{3+}$ ions.}
For the cavity transmission experiments, the Ti:Sapphire laser (M Squared SolsTiS) was continuously scanned at a rate of 10 MHz per second. The high-sensitivity CCD camera in the spectrometer (Princeton Instruments PIXIS) registers the transmitted photon counts intermittently at an adjustable frame rate (frames per second (fps)) with an exposure time 0.01 s for each frame. Therefore one exposure corresponds to a spectral width of 10 MHz$\times$0.01=100 kHz scanned by the laser, which is equal to the homogeneous linewidth of the 0.2\% doped sample. The long term drift of the laser is 10 MHz per hour, so the drift during each exposure should be inconsequential. Each data point in Fig.~\ref{f4}(b,c) represents the photon count collected in one camera exposure, corresponding to the signal contributed by the ions within one homogeneous linewidth. The frame rate was 0.1 fps for the coarser scan in Fig.~\ref{f4}(b), corresponding to a spectral interval of 100 MHz between two adjacent data points. The frame rate was 8.2 fps for the fine scan in Fig.~\ref{f4}(c), with a spectral interval $\sim1$ MHz. Each data point was obtained with one scan. Several scans at different spectral regions were performed and stitched together to cover the entire bandwidth in Fig.~\ref{f4}(b).

\noindent {\bf Estimation of the branching ratio.}
The measured optical depth of a 15-$\mu$m-long nano-beam resonator at 3.8 K was $d$=0.1, from which we deduce an oscillator strength of $f=6.5\times10^{-7}$ and a spontaneous emission rate of this transition to be $\gamma_{\rm 883}=1/\tau_{\rm 883}$=1/5.6 ms \cite{Mcauslan}. With a measured bulk medium lifetime $\tau_0$=250 $\mu s$, the branching ratio was thus estimated to be $\tau_0/\tau_{\rm 883}\approx4.5\%$.

\section*{Acknowledgements}
We gratefully acknowledge the contributions of Alexander E. Hartz. This work was funded by California Institute of Technology (Caltech). Equipment funding was also provided by IQIM, an NSF Physics Frontiers Center with support of the Moore Foundation. The device nanofabrication was performed in the Kavli Nanoscience Institute at Caltech.

\section*{Author contributions}
A.F. and T.Z. conceived the experiments. T.Z. and E.M performed the simulations and T.Z. fabricated the devices. T.Z., J.M.K. performed the measurements. T.Z. and J.M.K. analyzed the data. T.Z. and A.F. wrote the manuscript with input from all authors.

\section*{Competing financial interests} The authors declare no competing financial interests.

\clearpage

\newcommand{\beginsupplement}{%
        \setcounter{table}{0}
        \renewcommand{\thetable}{S\arabic{table}}%
        \setcounter{figure}{0}
        \renewcommand{\thefigure}{S\arabic{figure}}%
     }

\onecolumngrid

      \beginsupplement

\section{Supplementary Information}
      
\subsection{S1. Fabrication and characterization of arrays of YSO nano-beam resonators}
\noindent The YSO nano-resonators were fabricated in batch with careful focused ion beam (FIB) alignment and drift compensation. Figure~\ref{fS2} (a) shows a SEM image of an array of six Nd:YSO nano-resonators. All of the devices have resonance modes near the designed 883 nm wavelength, shown in the transmission spectrum in Fig.~\ref{fS2}(b). The color of each spectrum maps to that of the device in Fig.~\ref{fS2} (a). Measured quality factors in this batch range from 1,500 to 4,400. The spread of resonance wavelengths is about 10 nm, indicating the robustness of this fabrication process.

\begin{figure}[htb]
\includegraphics[width=1\textwidth]{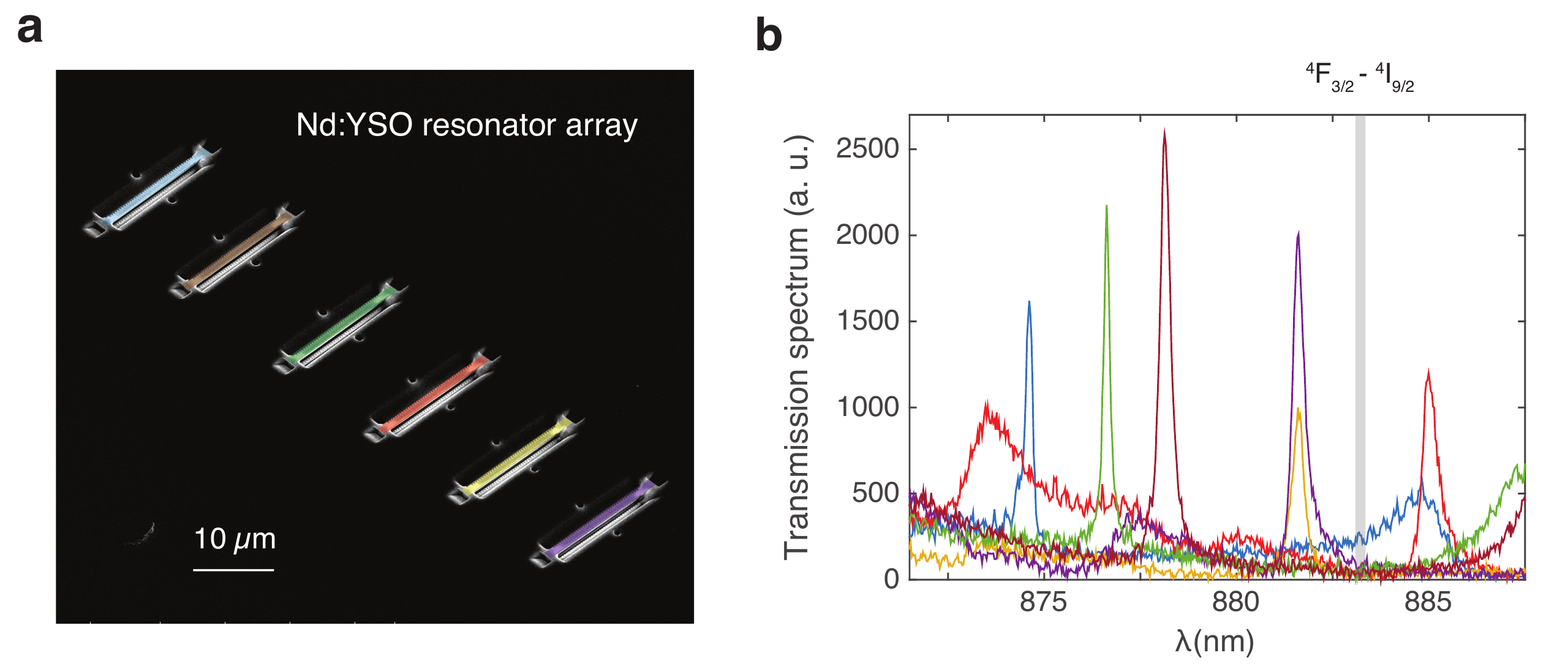}
\caption{{\bf An array of nano-beam resonators fabricated in Nd:YSO.} {\bf a,} Scanning electron microscope image of 6 devices. {\bf b,} Measured resonance modes (color matched to the corresponding device) near the 883 nm transition of Nd:YSO (grey line).}\label{fS2}
\end{figure}

\subsection{S2. Measurements of inhomogeneous linewidths for cavity-coupled ions} 
\noindent The inhomogeneous linewidth for the ions coupled to the 0.2\% nano-cavity was measured to be $\Delta_{\rm inhom}$=16.0 GHz from the dipole-induced transparency signal in Fig.~4(b). For the 0.003\% cavity, ion density was too low for a similar measurement of the inhomogeneously broadened distribution. Instead, the linewidth of photoluminescence (PL) from the cavity was measured with a high resolution spectrometer, and $\Delta_{\rm inhom}$=5.9 GHz was estimated by deconvolving the PL signal with the minimally resolvable linewidth of the spectrometer. For both doping levels, the same inhomogeneous linewidths were measured from the bulk via absorption spectroscopy. The agreement between the inhomogeneous linewidth of the cavity-coupled Nd ions and the bulk confirms the excellent spectral stability of REIs when embedded in nanophotonic resonators.

\subsection{S3. Requirement on the inhomogeneous linewidths for scalable QLMIs}
\noindent We consider a network of QLMIs each being a nano-cavity coupled to ensembles of emitters with inhomogeneous linewidth $\Delta_{\rm inhom}$.  Efficient QLMIs require the emitters to emit photons dominantly into the cavity mode. The cavity photons in a single spatial mode could then be efficiently coupled to waveguides or fibres for routing to other QLMIs operating at the same frequency. The probability of an emitter to emit a photon into the cavity mode is $\beta F/(1+(F-1)\beta)$, where $\beta$ is the branching ratio of the dipole transition, and $F$ is the Purcell factor in Eq.~\ref{eq2}. Assuming 99\% of the dipole emission into the cavity and typical branching ratio of $\beta\approx10\%$, the required Purcell factor should be $\sim$1,000. Considering a photonic crystal nano-cavity with a small mode volume of $1(\lambda/n)^3$, $F\sim1,000$ corresponds to a quality factor $Q\approx1\times10^4$ and a cavity linewidth $\kappa\approx$30 GHz (for the 883 nm transition). Thus, for scalability, the emitters and cavities need to be aligned within $\approx$10 GHz (order of magnitude). This limits the inhomogeneous broadening of the dipole ensembles to be $<$10 GHz for implementing robust and scalable QLMIs. This condition is satisfied by most REI transitions.

\subsection{S4. Calculation of ensemble averaged Purcell enhancement factor}
\noindent The spontaneous emission rate of a dipole coupled to a nano-resonator is enhanced relative to the bulk medium, by the factor $\sim1+\beta F$ \cite{Faraon}, where $\beta$ is the branching ratio of the transition, and $F$ is given by \cite{Purcell},
\begin{equation}
F=F_{\rm cav}\left(\frac{{\rm E(r)\cdot{\rm \mu}}}{{\rm |E_{max}||{\rm \mu}|}}\right)^2 \frac{1}{1+4Q^2(\lambda/\lambda_{\rm cav}-1)^2}
\label{eq1}
\end{equation}
where $\mu$ is the dipole moment, $\rm E(r)$ is the local electric field at the emitter location $\rm r$, $\lambda_{\rm cav}$ is the cavity resonant wavelength, $\lambda$ is the emitter wavelength, and $|\rm E_{max}|$ is the maximum electric field in the resonator. For a dipole that is resonant with the cavity and ideally positioned and oriented with respect to the maximum cavity field,
\begin{equation}
F_{\rm cav}=\frac{3}{4\pi^2}\left(\frac{\lambda_{\rm cav}}{n}\right)^3\frac{Q}{V_{\rm mode}}.
\label{eq2}
\end{equation}
We consider an ensemble of Nd ions uniformly distributed inside the YSO cavity. The enhancement of the emission from the ensemble can be estimated by averaging $F$ (Eq.~\ref{eq1}) over the entire population of Nd ions in the cavity. Based on the 3 dimensional field profile in Fig.~1(b), the mode volume 1.65$(\lambda/n)^3$ and $Q$=4,400, we numerically calculate this averaged Purcell factor to be 45 when the cavity is resonant with the transition. 

If the emission rate for uncoupled Nd ions is $1/\tau_{\rm 0}=1/\tau_{\rm 883}+1/\tau_{\rm other}$, in the coupled case the rate becomes $1/\tau_{\rm c}=(1+F)/\tau_{\rm 883}+1/\tau_{\rm other}$, where $1/\tau_{\rm 883}$ and $1/\tau_{\rm other}$ are the spontaneous emission rates into the 883 nm transition and other 4f-4f transitions, respectively. The Purcell factor is then experimentally extracted as $F=(\tau_{\rm 0}/\tau_{\rm c}-1)/\beta$, where $\beta$ is the branching ratio of the 883 nm line. Based on the measured branching ratio $\beta$= 4.5\%, the observed change in lifetimes leads to an ensemble averaged Purcell factor $\sim42$, which matches well with the calculated value. Furthermore, the averaged value of 42 means the expected Purcell enhancement for an ideally positioned and oriented Nd dipole is $F_{\rm cav}\approx200$.

\begin{figure}[htb]
\includegraphics[width=0.6\textwidth]{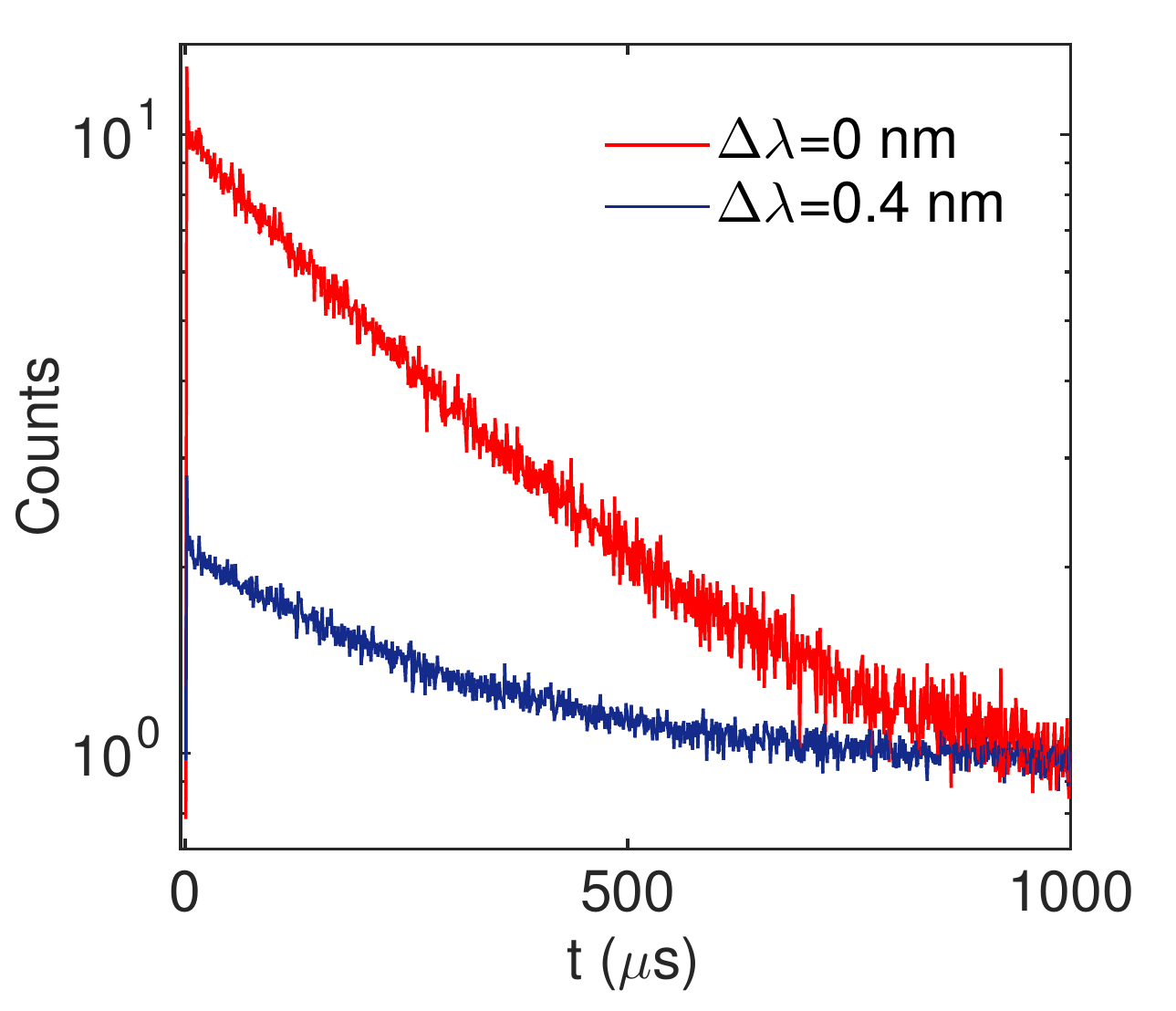}
\caption{{\bf Lifetime measurements for a 0.003\% doped Nd:YSO resonator.} Lifetime changes from 289 $\mu$s when the cavity of $Q$=1,500 is detuned by $\Delta$=0.4 nm, to 200 $\mu$s when the cavity is resonant with the ions.}
\label{fS1}
\end{figure}

\subsection{S5. Purcell enhancement in 0.003\% doped Nd:YSO nano-resonators}
\noindent A 0.003\% doped Nd:YSO nano-resonator was fabricated, measuring a resonance mode at 879 nm with quality factor $Q$=1,500. Spontaneous emission rate enhancement in this cavity was estimated from lifetime measurements in the same way as for the 0.2\% cavity. A change of lifetime from 289 $\mu s$ when the cavity is detuned by $\Delta=$0.4 nm, to 195 $\mu s$ at resonance ((Figure~\ref{fS2}) gives rise to an ensemble averaged Purcell factor F$\sim10$. Note that a longer $T_1=290 \mu s$ in the low density Nd:YSO bulk sample ($T_1=300 \mu s$ reported in \cite{Usmani}) yields a slightly larger branching ratio $\beta\sim5.4\%$. The longer $T_1$ is most likely due to weaker dipole-dipole interactions in 0.003\% doped sample.

\subsection{S6. Spectral diffusion of Nd$^{3+}$ ions coupled to the nano-cavities }
\noindent The dynamic coherence properties of the cavity-coupled Nd ions were investigated by three-pulse photon echoes (($\pi/2$ - $\pi/2$ - $\pi/2$)) (Figure~\ref{fS3}(a) )that gives information about the spectral diffusion on time scales up to $T_1$ \cite{Perrot} . The third pulse, delayed by a time $T_{\rm w}$ after second pulse, is diffracted on the spectral grating from the first two pulses and produces an echo. Spectral diffusion - frequency shifts of the optical transition due to the fluctuating rare earth environment - gradually erases the grating during $T_{\rm w}$, and causes faster echo decays thus broadening of the effective linewidth $\Gamma_{\rm eff}$. Linearly increasing $\Gamma_{\rm eff}=\Gamma_{\rm h}+RT_{\rm w}$ at a rate of  $R$ = 380 Hz $\mu$s$^{-1}$ was measured for the 0.003\% cavity and 6.1 kHz $\mu$s$^{-1}$ for the 0.2\% cavity (Figure~\ref{fS3}(b,c)). Higher spectral diffusion is expected for higher doping because of stronger dipole-dipole interaction between Nd ions. Nevertheless, the measured linewidth broadening is much smaller than our Rabi frequency ($\sim$6 MHz in Fig.~3(c)). This indicates the coupled ions, either singles or ensembles, can be optical addressed repeatedly up to 10s of $\mu$s, which is desirable for optical quantum information processing. 
\begin{figure}[hbt]
\includegraphics[width=1\textwidth]{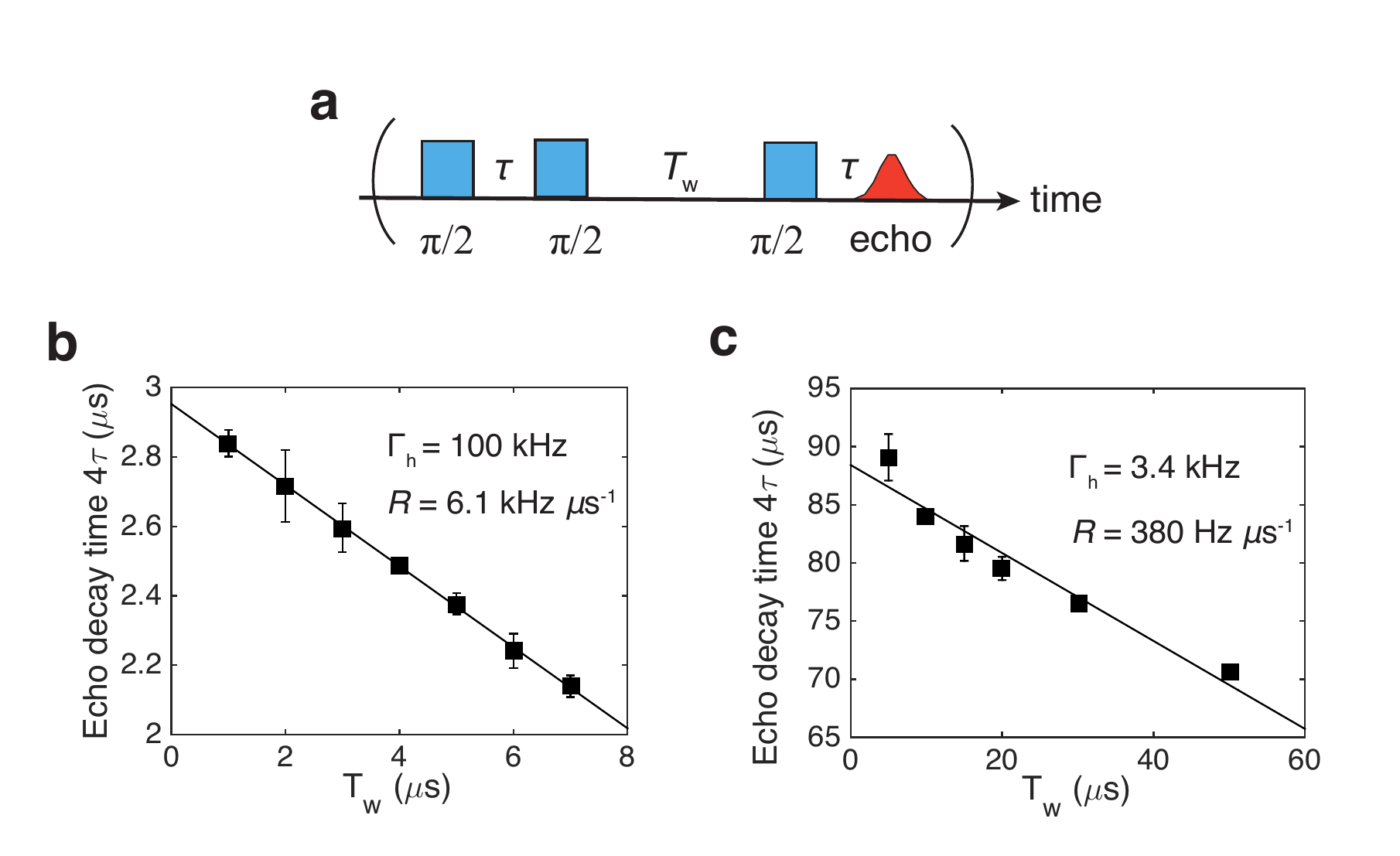}
\caption{{\bf Spectral diffusion of cavity-coupled ions} {\bf a,} Three-pulse photon echo sequence ($\pi/2$ - $\pi/2$ - $\pi/2$). {\bf b-c,} $T_{\rm w}$ dependent broadening of the effective linewidths of ions coupled to the 0.2\% ({\bf b}) and 0.003\% ({\bf c}) doped cavities, measured via three-pulse photon echoes. Linear fits indicate the spectral diffusion rates $R$ for both doping concentrations.}
\label{fS3}
\end{figure}

\subsection{S7. REI-controlled cavity transmission and saturation of the coupled ions}
\noindent Figure~\ref{fS4} plots the on resonance transmission at zero detuning as a function of the average photon number in the cavity $\langle n_{\rm cav} \rangle$. $\langle n_{\rm cav} \rangle$ was estimated from the input probe laser power P$_{\rm in}$ (measured after the objective), the coupler efficiency $\eta$, and cavity coupling rate ($\kappa/2$) as $\langle n_{\rm cav} \rangle=\eta P_{\rm in} / \kappa \hbar \omega$. Black curve is the theoretical calculation using the Quantum toolbox \cite{Tan}, which shows close agreement with the experiment.
\begin{figure}[hbt]
\includegraphics[width=0.6\textwidth]{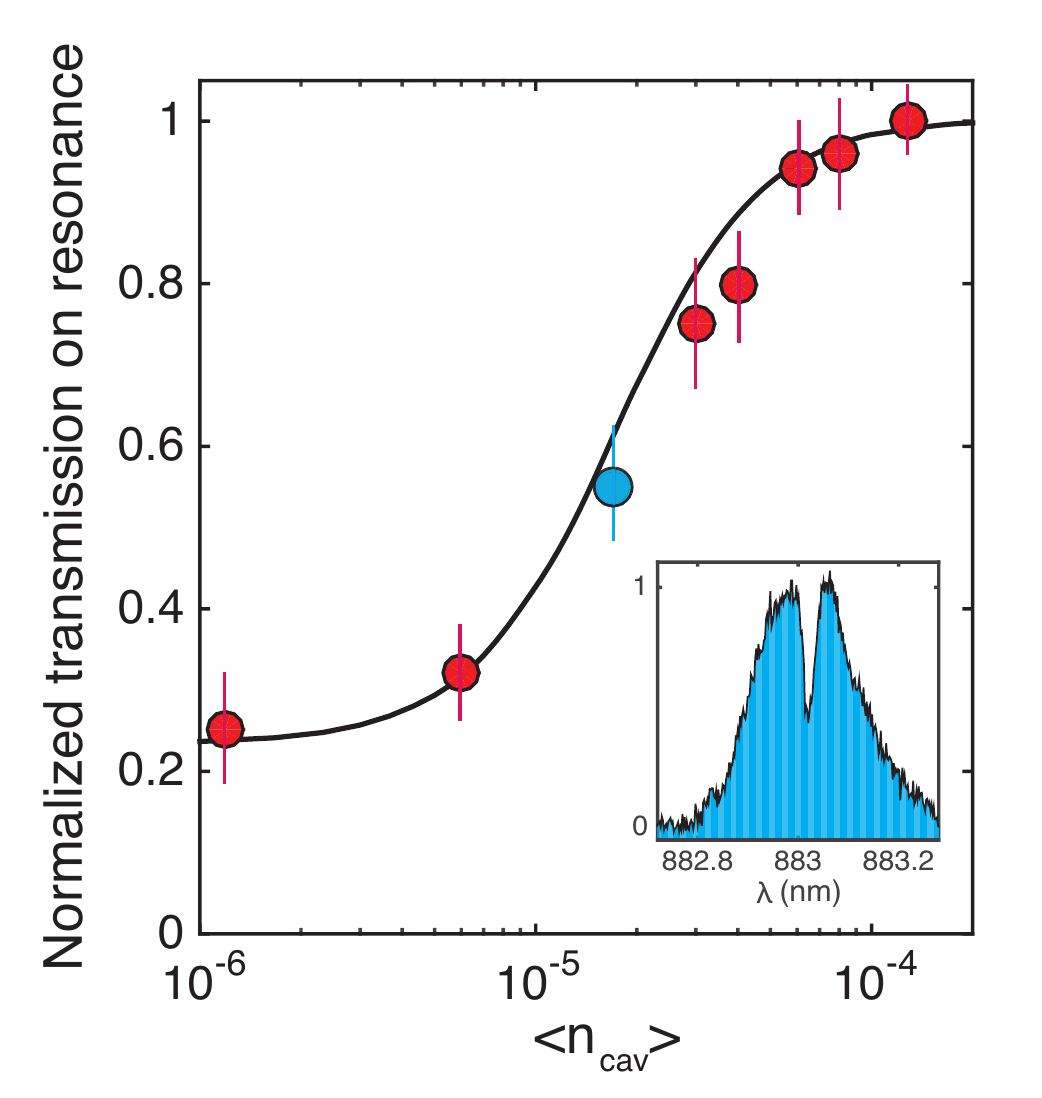}
\caption{{\bf Controlled cavity transmission versus intra-cavity photon number for probe laser at zero detuning.} The cavity transmission is normalized by empty-cavity transmission at the zero detuning. Saturation occurs at $\langle n_{\rm cav} \rangle \approx2\times10^{-5}$. The inset shows a transmission spectrum at onset of saturation with a normalized transmission at the dip $\approx$54\%.  }
\label{fS4}
\end{figure}

\subsection{S8. Towards detection and control of single REI ions coupled to the nano-cavity}
\noindent The measurement of $N$($\Delta \lambda$) indicate that this system can be used to detect and control a single ion coupled to the cavity - a key ingredient for realizing quantum networks interconnecting multiple quantum bits encoded in individual REI ions. In the 0.003\% low density devices, we estimated a peak ion density of $N$=0.07 per $\Gamma{\rm _h}$=1/$\pi T_2$=3.1 kHz. Correspondingly, the single ion cooperativity of $\eta=$1.6 can be attained with the same cavity $Q$=4,400, $V=1.65(\lambda/n)^3$, and $g=2\pi\times$10 MHz (typical for REI transitions and we assume the ion is positioned at maximum cavity field). Simulation using Quantum Optics Toolbox \cite{Tan} yields a transmission dip $>$80\% due to a single Nd ion, as shown in Fig.~\ref{fS5}. The main technical challenge to detect single ion in this cavity system is the requirement of a highly stabilized laser, with linewidth $<$1 kHz and minimal long term drift, for scanning the single ion spectrum, which should be attainable with state of the art laser spectroscopy technology.

\begin{figure}[htb]
\includegraphics[width=0.55\textwidth]{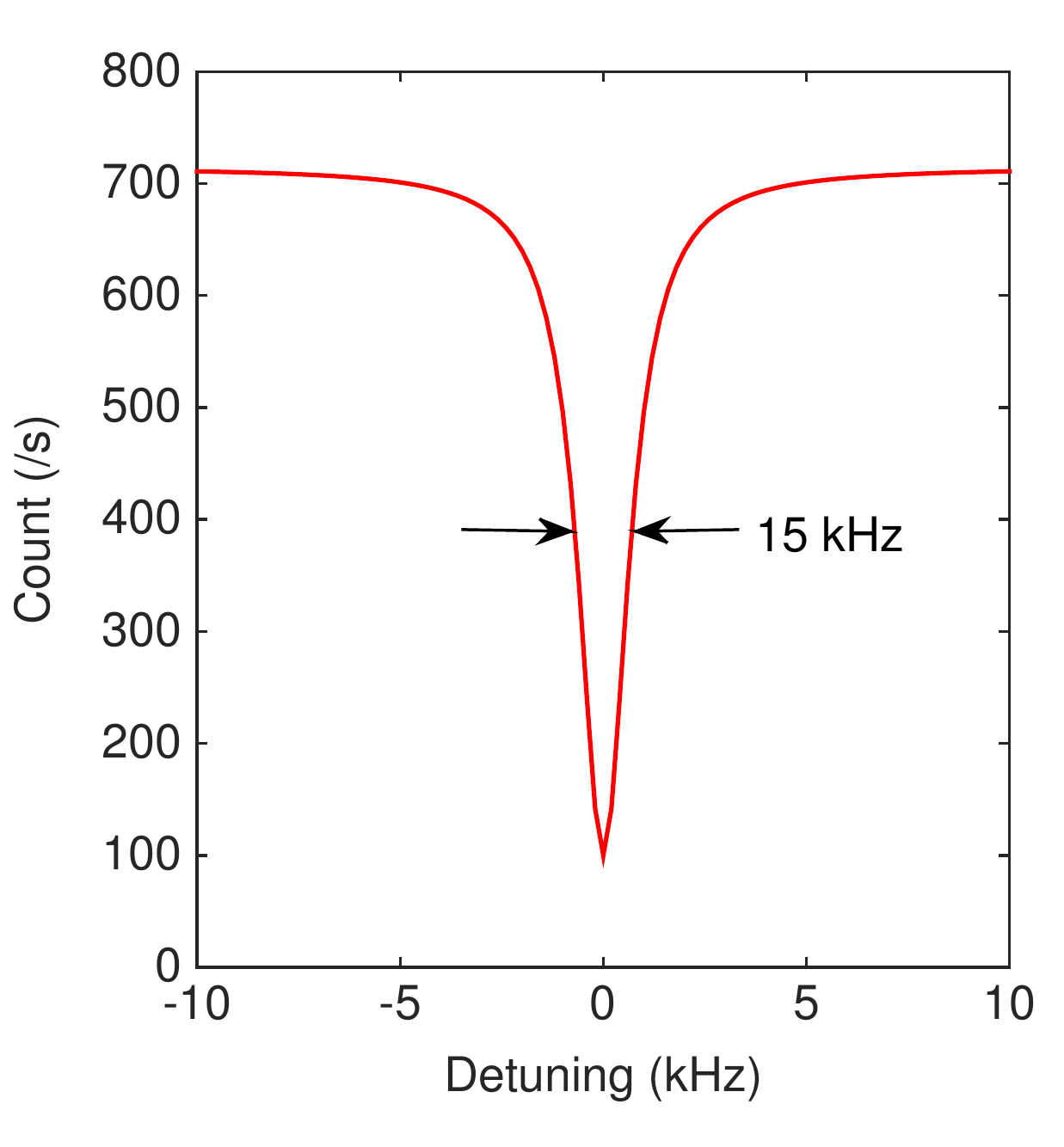}
\caption{{\bf Cavity transmission when coupled to a single Nd$^{+3}$ ion.} The simulation uses Quantum Optics Toolbox \cite{Tan} with parameters $Q$=4,400, $\Gamma_{\rm h}$=3.1 kHz, and g$=2\pi\times$10 MHz. The transmission dip has a full-width at half-maximum (FWHM) of $\approx$15 kHz.}
\label{fS5}
\end{figure}

\end{document}